\documentclass[aps, prl, twocolumn, amsmath, graphicx, latexsym, asmsymb, amsfonts]{revtex4}

\usepackage{epsfig}

\def\com#1{{\tt [\hskip.5cm #1 \hskip.5cm ]}}

\DeclareMathOperator{\tr}{tr}
\DeclareMathOperator{\Id}{ {\bf 1}}

\begin{document}

\title{Quantum Information Processing in Disordered and Complex Quantum Systems}
\author{Aditi Sen(De)\(^{1,2}\), Ujjwal Sen\(^{1,2}\), Veronica Ahufinger\(^3\),  Hans J. Briegel\(^{4,5}\), Anna Sanpera\(^{2,6,*}\),
Maciej Lewenstein$^{1,2,*}$}

\affiliation{\(^1\)ICFO-Institut de Ci\`encies Fot\`oniques, 
E-08034 Barcelona, Spain \\
$^2$Institut f\"ur Theoretische Physik,
Universit\"at Hannover, D-30167 Hannover,
Germany\\
\(^3\)Grup d'Optica, Universitat Aut\`onoma de Barcelona, E-08193 Bellaterra, Spain\\
\(^4\)Institut f\"ur Quantenoptik und Quanteninformation, \"Osterreichische Akademie der Wissenschaften, A-6020 Innsbruck, Austria\\
\(^5\)Institut f\"ur Theoretische Physik, Universit\"at Innsbruck, Technikerstra{\ss}e 25, A-6020 Innsbruck, Austria\\
\(^6\)Grup de F\'isica Te\`orica, Universitat Aut\`onoma de Barcelona, E-08193 Bellaterra, Spain
}



\begin{abstract}
We investigate quantum information processing and manipulations in disordered systems of ultracold atoms and trapped ions. First,
we demonstrate generation of entanglement and {\sl local} realization of quantum gates in a quantum spin glass system.  Entanglement in
such systems attains significantly high values, after quenched averaging, and has a stable positive value for arbitrary times.
Complex systems with long range interactions, 
such as ion chains or dipolar atomic gases, can be modeled by  
neural network Hamiltonians.
In such systems, we find the characteristic time of persistence of
quenched averaged entanglement, and also find the time of its revival.

%
%
\end{abstract}
\maketitle

\def\com#1{{\tt [\hskip.5cm #1 \hskip.5cm ]}}


Successful implementations of quantum information processing (QIP) in atomic, molecular, or solid state systems
typically demand  very rigorous
control of such systems \cite{general}. 
This concerns both few qubit systems such as the Cirac-Zoller computer 
\cite{cirac} with ions or photons \cite{experiments}, as well as 
atomic gases 
in optical lattices \cite{jaksch}.
Despite a lot of progress, the demanded control in such systems  
is nowadays very hard to achieve
\cite{daley}. Recently QIP in systems
with a limited knowledge of the parameters has also been proposed \cite{ripoll}.

At the first sight, what we propose here sounds like  {\sl contradictio in adjecto}:
QIP in quenched disordered or complex, {\sl ergo} hardly controllable, systems. However,
as we have recently shown, one can create {\sl controlled} disorder in atomic gases in optical
lattices and study, in an unconventional way, 
Anderson and Bose glasses in a Bose gas \cite{damski},
or spin glasses with short range interactions  
in  Fermi-Bose, or Bose-Bose mixtures \cite{sanpera}.
Using linear chains of trapped ions \cite{porras}, or dipolar atomic gases \cite{Baranov}, it
is possible to realize complex spin systems with long-range interactions
that may serve as model for classical and quantum neural networks 
\cite{marisa}.

Disordered systems offer at least two possible advantages for QIP. First, 
they have typically a large number of different metastable (free) 
energy minima, as it happens in spin glasses (SG) \cite{parisi}. 
Such states might be used to store information distributed over the whole
system, similarly to neural network (NN) models \cite{amit}. 
The information is thus naturally stored in a redundant way, 
like in error correcting schemes \cite{shor}. 
Second, in disordered systems with long range interactions, 
the stored information is robust: metastable states have quite large
basins of attraction in the thermodynamical sense.

We address here the simplest fundamental questions concerning QIP in disordered or complex
systems: (i) 
Can one generate
entanglement in such systems that would survive quenched averaging over long times? (ii)  Can one realize quantum gates with reasonable fidelity?
 Here we answer both questions affirmatively considering both
short and long range disordered systems.

First, we consider a short range disorder Ising Hamiltonian,
the so-called Edwards-Anderson (E-A) model of spin glasses which
can be straightforwardly implemented using atomic Bose-Fermi, or
Bose-Bose mixtures in optical lattices\cite{bofe,sanpera}.
We address the generation and evolution 
of nearest neighbor (nn) entanglement in this model. 
In the 
short range Ising model without disorder,
it is possible to create cluster and graph states (i.e. entanglement) 
starting from an appropriate initial product state \cite{briegel}.  
Here we show that, while the disorder averaged density matrix 
of two neighboring spins remains always separable,
the disorder averaged entanglement 
(quantified by logarithmic negativity \cite{VidalWerner})
converges with time to a finite value. 
The generation of entanglement \cite{briegel} as well as 
its evolution for arbitrary times in an Ising model without disorder
but with long-range interactions, 
has also been addressed in Ref. \cite{briegelnew}. There
it was suggested the possibility of applying  
similar ideas
to  disordered systems. We show
also that the quantum single-qubit Hadamard gate,
can be realized in such system with significant (disorder averaged) fidelity.

Secondly, we consider complex systems with long range ($1/r^3$, or $1/r^2$) interactions, that can
be realized for instance, in linear ion traps, using either local magnetic fields, as proposed by 
Wunderlich and coworkers\cite{christoph}, or by appropriately designed laser
excitations \cite{porras}. The corresponding Hamiltonian
can be mapped into an Ising
Neural Network (NN) model with weighted patterns \cite{amit}. Those patterns can be used as qubit systems, with the information distributed over the chain. 
One can also include external parallel, or transverse fields
in the model.
We show that in such  system, it is
possible to generate long range bipartite entanglement that undergoes a series of collapses and revivals \cite{Eberly},
 whose times are found analytically.
Finally we study also bipartite and tripartite entanglement dynamics in an infinite range Ising model without disorder.

Let us start with the Edwards-Anderson spin glass model described by
\begin{equation}
\label{eqspinglass}
H_{E-A} = - \frac{1}{4} \sum_{\left\langle ij\right\rangle }
J_{ij} \sigma^z_i \sigma^z_j.
\end{equation}
Here \(\sigma^z_k\) denotes the Pauli 
operator at the \(k\)th site, and
\(J_{ij}\)'s describe nn couplings for
an arbitrary lattice. In the E-A model these couplings 
are given by independent Gaussian variables with mean \(J\) 
and variance \(\sigma^2\).
Starting from a pure product state of the form
$|\Psi\rangle=\prod_{i}|+\rangle_i$,
where \(\left|\pm\right\rangle = (\left|0\right\rangle \pm \left|1\right\rangle)/\sqrt{2}\)  \cite{briegelnew},
 we evaluate the entanglement after a finite time, where 
the density matrix is given by \(\rho(t,\{J_{ij}\}) = \exp\{-iH_{E-A}t\}|\Psi\rangle\langle \Psi|\exp\{+iH_{E-A}t\}\). 
The reduced density matrix for 
a nn pair is obtained by tracing over all other sites.
For instance, the reduced density matrix for a  2D square lattice
is given by
\begin{eqnarray}
&&\varrho_{12}(t, \{J_{ij}\})=\frac{1}{4} \Id \otimes \Id + \frac{1}{4} \Big[
e^{iJ_{12}t/2}\\
&&\Big\{\cos(J_{24}t/2) \cos(J_{26}t/2)\cos(J_{28}t/2) |00\rangle \langle 01| 
\nonumber\\
&+& \cos(J_{13}t/2) \cos(J_{15}t/2)\cos(J_{17}t/2) 
|00\rangle \langle 10| \Big\}\nonumber\\
&+&e^{-iJ_{12}t/2} 
\Big\{\cos(J_{13}t/2 \cos(J_{15}t/2) \cos(J_{17}t/2) 
|01\rangle \langle 11| \nonumber\\
&+& \cos(J_{24}t/2) \cos(J_{26}t/2) \cos(J_{28} t/2)
 |10\rangle \langle 11|    \Big\}\nonumber\\
&+& \cos(J_{13}t/2) \cos(J_{15}t/2) \cos(J_{17}t/2)
 \cos(J_{24}t/2)\nonumber\\
&\times& \cos(J_{26}t/2) \cos(J_{28} t/2)
 \Big\{|00\rangle \langle 11|  + |01\rangle \langle 10| \Big\}
+ \mbox{h.c.}\nonumber
\Big],
\end{eqnarray}
where $\Id$ is the identity operator and the indices \(3 \ldots 8\) 
enumerate the six neighbors of 1 and 2. A 
similar
expression can be 
obtained for the 1D lattice. In both cases,
the averaging of the reduced state over \(J_{ij}\)'s (equivalent 
to reducing  the average \(\varrho_{12}(t, \{J_{ij}\})\)) is \emph{separable}. 
Note, however, that as always in physics of disordered systems, if 
we are interested in typical values of physical quantities such as free energy, entanglement, etc., we are obliged to 
perform a ''quenched'' average, i.e. first calculate the quantity of interest and then average 
\cite{parisi} (see also \cite{ekhaneo1,ekhaneo2}).


To study entanglement, we use the logarithmic negativity (LN) \cite{VidalWerner}. The
LN of a bipartite state \(\rho_{AB}\) is defined as
\(E_{LN}(\rho_{AB})= \log_2 \|\rho_{AB}^{T_{A}}\|_1\),
where \(\|.\|_1\) is the trace norm, and \(\rho_{AB}^{T_{A}}\) denotes the partial transpose of
\(\rho_{AB}\) with respect to the \(A\)-part \cite{Peres_Horodecki}.
Note that $\rho_{ij}(t)$ acts on \(\mathbb{C}^2 \otimes \mathbb{C}^2\).
Consequently, a positive value of the LN implies that the state is
entangled and distillable \cite{Peres_Horodecki, Horodecki_distillable}, while
\(E_{LN} =0\) implies separability
\cite{Peres_Horodecki}.

\begin{figure}[h]
\begin{center}
\includegraphics[width=0.9\linewidth]{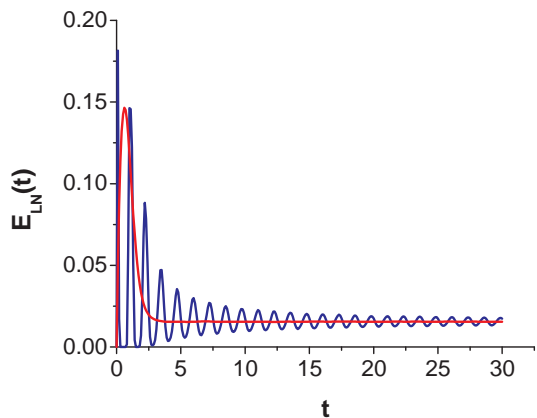}
\caption{
Temporal behavior of nn averaged entanglement in a 2D spin glass model, 
starting from $\Pi_i\left| + \right\rangle_i$.
For a model with frustration 
($J=0$), \(E_{LN}(t)\) converges quickly to a constant value (red curve). 
For a non-frustrated case (\(J=+5\)),  \(E_{LN}(t)\) 
exhibits damped oscillations (blue curve), converging to 
the same value \(\approx 0.0154\), as reached in the frustrated case. 
Standard deviation for \(t\rightarrow \infty\)
is \(\approx 0.0704\). 
It is interesting to note that the dynamical behaviour of \(E_{LN}\) 
depends on $J$, although at large times, they all converge to the same
value.The same behavior is encountered in the 1D case, 
even though there is no frustration in that case.
}
\label{fig_2dlattice}
\end{center}
\end{figure}

The entanglement in the spin glass model turns out to be 
an even function of the couplings.
The temporal behavior of \(E_{LN}(t)\) in a 2D square lattice is 
shown in Fig. \ref{fig_2dlattice} for two different
cases of disorder: with frustration and without it. For \(J = 0\), \(\sigma^2 =1\), 
the system has randomly ferro- ($J>0$) 
and antiferro-magnetic ($J<0$) interactions 
and is strongly frustrated; \(E_{LN}(t)\) is rapidly damped 
to a constant, and does not show any oscillations. This behaviour 
differs from the non-frustrated case \(J = 5\), \(\sigma^2 = 1\),
when \(E_{LN}(t)\) exhibits oscillations with frequencies \(\sim  1/J\).
For short range interactions,  the next-nearest neighbor 
entanglement vanishes, even \emph{before} the averaging, 
for both 1D and 2D. 
To understand why entanglement converges in time to the same finite value 
in both the frustated and non-frustated cases, 
notice that as long as the distributions \(J_{ij}\)'s are sufficiently well-behaved, 
\(J_{ij}t/2\) corresponds to a uniform distribution over \([0,2\pi]\) for large enough \(t\).

We have calculated the nn entanglement for the following lattice configurations:
1D chain, 2D honey-comb lattice, 2D square,  3D cube, where 
any given pair of neighboring lattice sites has \(d=2,4,6,10\) 
neighbors respectively. For time large enough, our numerics
reveal that bipartite entanglement decays exponentially
with the number of neighbors. Such behaviour can be reproduced 
analytically by considering the volume of the set of separable states 
(see e.g. \cite{KarolMaciek}), giving an upper bound on nn entanglement
that depends exponentially on $d$. Some algebra shows
that if the state $\rho_{ij}(t)$ is entangled, then
\(\sum_{i=1}^{d} \phi_{i}^2 < (3 - 4R^2)/2\), where the 
\(\phi_{i}= J_{ij} t/2\)'s are state parameters varying from 0 to \(2 \pi\),
and \(R\) is the radius of the separable ball in the $d$-dimensional space. 
The volume of this hypersphere is \({\cal V}_d= {\cal S}_d \left(\frac{3-4R^2}{2}\right)^\frac{d}{2}/d\),
where \({\cal S}_{d} = 2\pi^{\frac{d}{2}}/\Gamma(d/2)\).
Due to the periodicity involved implicitly in $\rho_{ij}(t)$, 
there are \(2^d -1\) such hyperspheres.
Considering all states in this volume to have unit entanglement, the average entanglement at long times is
\({\cal E}_d = {\cal V}_d(2^d -1)/(2\pi)^d\).
As an example, for the case of the 2D lattice (for
which \(d=6\)), at long times, the actual entanglement is \(\approx 0.0154\), while \({\cal E}_6 \approx 0.0221\).
Although the bipartite entanglement vanishes with increasing number of neighbors, one can expect the multipartite
entanglement to be non vanishing due to the fact that the volume of separable states is ``super-doubly-exponentially  small''
with increasing number of parties \cite{dobol-chhoto}.

We show now that spin glasses allows also 
to implement quantum gates.  We focus on
the Hadamard gate, 
which transforms the computational basis into a complementary basis:
\(\left|0\right\rangle \rightarrow  \left|+\right\rangle\) and
\(\left|1\right\rangle \rightarrow  \left|-\right\rangle\).
To implement the Hadamard gate, 
assume that the computation is performed in a spin lattice, 
and the particles 1 and 2 are a part of it.
We assume that at a certain time, particle 1
is in an arbitrary state \(a|0\rangle + b|1\rangle\), 
where \(|a|^2 + |b|^2 =1 \), and we let
system evolve according to the Hamiltonian \(H_{E-A}\) 
for a suitable duration of time, before performing
measurement on particle 1 (in a suitable basis).
For \(J = 5, \sigma^2 =1 \), particle 2 attains the Hadamard rotated 
state \(a|+\rangle + b|-\rangle\), with quenched averaged
fidelity greater
than \(0.85\). One can increase such fidelity by increasing the 
number of spins, and employing assisted measurements.
Note, that if we try to prepare the Hadamard rotated state 
using the classical information obtained only from the
 measurement of particle
 \(1\), the fidelity is only \(2/3\) \cite{classicalfidelity}.

Let us now move to a \emph{long-range} interactions spin Ising model,
described by  the Hamiltonian
\(H_{lr} = \frac{1}{N}\sum_{i,j} J_{ij}\sigma_i^z \sigma_j^z \),
where \(N\) is the total number of spins. Such models can be realized with trapped ions \cite{marisa}, where
\(J_{ij} = \sum_\mu \xi^i_{\mu} \xi^j_{\mu}/\lambda_{\mu}^2\), with \(\xi^{i}_{\mu}\) (\(\lambda_\mu\))
describing the phonon eigen-modes (eigen-frequencies).
Here we consider two extreme cases. First, we take \(\lambda_1 = 1,  \xi_1^i = \mbox{constant}\)  \(\forall i\),
\(\lambda_{\mu} \rightarrow \infty\)  for
\(\mu \geq 2\), so that  the interactions are ordered, and the Hamiltonian is \(H_{lro} = \frac{1}{N} S^2\),
where \(S = \sum_{i=1}^{N}\sigma^z_i\).
Secondly, we consider the case when \(\lambda_\mu =1\) for all \(\mu\), when
the Hamiltonian
becomes
\(
H_{NN} = \frac{1}{N} \sum_{i,j =1}^{N} \sum_{\mu =1}^{p} \xi^{(i)}_\mu \xi^{(j)}_\mu \sigma_i^z \sigma_j^z
\).  This is  the Hopfield model of a \emph{neural network} with Hebbian couplings \cite{amit}.
Here \(p\) is the number of ``patterns" of the neural network, and the patterns are described
by random variables  \(\xi^{(i)}_\mu = \pm 1\), each with probability \(\frac{1}{2}\).
As in the case of short-range interactions, we take the initial state of the evolution as
\(|\Psi\rangle = \Pi_{i=1}^{N} |+\rangle_i\), and study the dynamics of entanglement 
for ordered and disordered Hamiltonians.
We provide an efficient method to analytically
compute the evolved state of any number of patterns and any number of spins.

\begin{figure}[h]
\begin{center}
\includegraphics[width=0.85\linewidth]{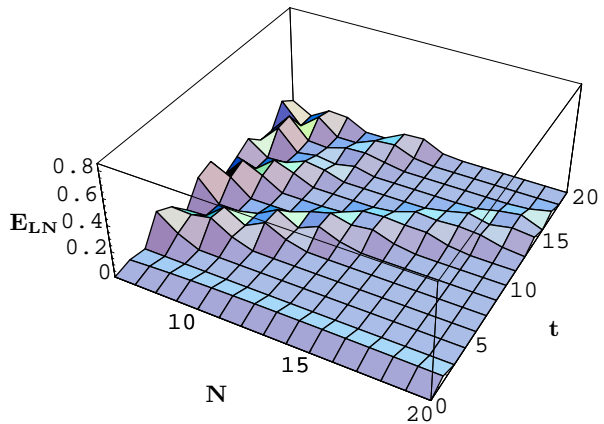}
\put(-150,15){\textbf{N}}
\put(-224,80){\(\textbf{E}_{\textbf{LN}}\)}
\put(-20,40){\textbf{t}}
\caption{Generation of entanglement of bipartite states \(\varrho^{lro}_{12}(t)\) with respect to time and number of spins. 
Collapses and revivals of the entanglement are clearly depicted.}
\label{fig_dui_N_t}
\end{center}
\end{figure}

Consider first the case of the Hamiltonian \(H_{lro}\).
We can write the evolution operator \(\exp(-i S^2 t/N)\) as
\(\int d\omega \exp( (i/N)\omega^2 + S \sqrt{t} (- 2 i/N)\omega)\), up to a
constant factor. Applying now this unitary to the initial state \(|\Psi\rangle\), we find  any two-party state
\(\varrho_{12}^{lro}(t) = \tr_{k\ne 1,2} \rho^{lro}(t)\) of such system and compute the entanglement quantified by the LN.
(This method can be also applied to find multipartite evolved states).
In Fig \ref{fig_dui_N_t},
we plot the entanglement (as quantified by LN) of \(\varrho^{lro}_{12}(t)\), with respect to time,
as well as \(N\).
The figure shows  revivals of
 bipartite entanglement, that occur on the time scale \(\tau_R \sim N\), and persist on the time
scale \(\tau_C \sim \sqrt{N}\) (collapse time).

As depicted in Fig. \ref{fig_dui_N_t}, there are large ranges of time, for which the bipartite state is separable.
Interestingly, this range of separability can be reduced,  considering entanglement of
the tripartite evolved state \(\rho_{123}^{lro}(t)\)
in a bipartite cut.
Although the interactions in \(H_{lro}\) are long-range, they are ordered,
so that
\(\rho_{12}^{lro}(t)\) and \(\rho_{123}^{lro}(t)\) takes a relatively simple
form.
Amazingly, the same method applies for \(H_{NN}\), where the
interactions are \emph{both} long-range and disordered.
%
%
Despite its increased complexity,  we can still
use the technique for the evolution operator \(\exp(-iH_{NN}t)\), that was used in the case of
\(H_{lro}\).
Specifically,
we replace in \(\exp(-iH_{NN}t)\), the operator \(\exp(-i S_\mu^2 t/N)\) by
\(\int d\omega_\mu \exp( (i/N)\omega_\mu^2 + S_\mu \sqrt{t} (- 2 i/N)\omega_\mu)\),
for every \(\mu\), where
\(S_\mu = \sum_{i=1}^{N} \xi^{(i)}_\mu \sigma_i^z\).
Applying this  operator to our initial state, we find
that
the  \(N\)-particle state at time \(t\) is
\begin{eqnarray}
\varrho^{NN}(t) = \int (\Pi_\mu dr_\mu ds_\mu{'}) \mbox{e}^{i\sum_\mu r_\mu s_\mu/N}
\nonumber \\
\Pi_{i=1}^{N} \Big[ \mbox{e}^{-2i\sqrt{t} \sum_\mu \xi_\mu^{(i)} s_\mu/N}
(|0\rangle \langle 0|)_i
\nonumber \\
+ \mbox{e}^{2i\sqrt{t} \sum_\mu \xi_\mu^{(i)} s_\mu/N} (|1\rangle \langle 1|)_i
\nonumber \\
+ \Big\{ \mbox{e}^{-2i\sqrt{t} \sum_\mu \xi_\mu^{(i)} r_\mu/N} (|0\rangle \langle 1|)_i + \mbox{h.c.} \Big\}
\Big],
\end{eqnarray}
where \(r_\mu = \omega_\mu + \omega_\mu{'}\),
\(s_\mu = \omega_\mu - \omega_\mu{'}\), with \(\mu = 1, \ldots, p\).
After tracing out all except particles 1 and 2 we obtain:
\begin{eqnarray}
&\varrho&^{NN}_{12}(t) = 1/4\Big\{(|00\rangle \langle 00| + |01\rangle \langle 01| +|10\rangle \langle 10| +|11\rangle \langle 11|) \nonumber \\
&+& \Big[ \mbox{e}^{- 4it \sum_\mu \xi_\mu ^{(1)} \xi_\mu ^{(2)}/N}
\Big( \Pi_{i \ne 1,2} \cos(4 t \sum_\mu \xi_\mu^{(i)} \xi_\mu^{(2)} /N)
 |00\rangle \langle 01| \nonumber \\
&+& \Pi_{i \ne 1,2} \cos(4 t \sum_\mu \xi_\mu^{(i)} \xi_\mu^{(1)} /N) 
|00\rangle \langle 10| \Big)\Big]\nonumber \\
&+&  \Pi_{i \ne 1,2} \cos(4 t \sum_\mu \xi_\mu^{(i)} (\xi_\mu^{(1)} + \xi_\mu^{(2)})/N)|00\rangle \langle 11|\nonumber\\
&+&  \Pi_{i \ne 1,2} \cos(4 t \sum_\mu \xi_\mu^{(i)} (\xi_\mu^{(1)} - \xi_\mu^{(2)})/N)|01\rangle \langle 10| \nonumber\\
&+& \Big[ \mbox{e}^{4it \sum_\mu \xi_\mu ^{(1)} \xi_\mu ^{(2)}/N}
\Big( \Pi_{i \ne 1,2} \cos(4 t \sum_\mu \xi_\mu^{(i)} \xi_\mu^{(1)}     |01\rangle \langle 11| \nonumber \\
&+& \Pi_{i \ne 1,2} \cos(4 t \sum_\mu \xi_\mu^{(i)} \xi_\mu^{(2)} /N) |10\rangle \langle 11| \Big)\Big]+  \mbox{h.c.} \Big]\Big\}. 
\end{eqnarray}
For \(N\) large, and \(t/N\) small, the above expression can be simplified using the 
fact that 
$\Pi_{i \ne 1,2} \cos(4 t \sum_\mu \xi_\mu^{(i)} \xi_\mu^{(2)} /N)
=\exp\Big[\sum_{i \ne 1,2}\log_e|\cos(4 t \sum_\mu \xi_\mu^{(i)} 
\xi_\mu^{(2)} /N)|\Big]
=\exp[ -( 8 t^2 /N^2) \sum _{i \ne 1,2} (\sum_\mu x^i_\mu)^2 ]$, 
where for all \(i\), \(x^i_\mu = +1\) or \(-1\) with probability \(1/2\) each. Therefore, for large \(N\) and small \(t/N\), we have
that \(
\Pi_{i \ne 1,2} \cos(4 t \sum_\mu \xi_\mu^{(i)} \xi_\mu^{(2)} /N)\) self-averages to the value
\( \exp[ -( 8 t^2 p/N)]
\),
so that after time \(t \sim \sqrt{N/p}\), all the off-diagonal elements of the state \(\varrho^{NN}_{12}(t)\) become vanishingly small.
Therefore, for the first time, nearest neighbor entanglement in the evolved state appears and  persists for  times of order
\(\tau_C \sim \sqrt{N/p}\).
However, there are repeated revivals in entanglement, with the period being \(\tau_R \simeq \pi N/2\) for odd \(p\), and
\(\tau_R \simeq \pi N\) for even \(p\). Note, that the period of revivals is independent
of the number of patterns in the  model (cf. \cite{ekhaneo2}).

Summarizing, we have studied disordered and complex spin systems with short-range and long range interactions
that can be realized with trapped atoms or ions. We have shown that
in both cases it is possible to generate quenched averaged entanglement over long times. In the case of
short range interactions, we considered Edwards-Anderson model in 1D and 2D square lattice.
We have shown that in such disordered system,
it is possible to implement also distinctly quantum single-qubit gates with high fidelity.
We have also demonstrated that it is possible to generate  entanglement in the spin system
 with long range interactions, corresponding to the  Hopfield neural network model. We have shown that
 in such case, entanglement exhibits a sequence of collapses and revivals.


We thank  I. Bloch, H.-P. B\"uchler, J. Eschner, M. Pons, L. Sanchez-Palencia,
J. Wehr and P. Zoller for fruitful discussions.
We acknowledge support from the Deutsche
Forschungsgemeinschaft (SFB 407, SPP1078 and SPP1116, 436POL), the RTN Cold Quantum Gases,
Ministerio de Ciencia y Tecnolog{\' i}a BFM-2002-02588, the Alexander von Humboldt Foundation, the EC Program
QUPRODIS, the ESF Program QUDEDIS, and EU IP SCALA.

\end{document}